\documentclass[12pt]{article}
\usepackage{amsthm,amssymb,amsmath}
\usepackage[american,british]{babel}

\renewcommand{\t}{{\operatorname{t}}}

 \newcommand{\res}{\operatorname{res}}

\newcommand{\Pf}{\operatorname{Pf}}
\newcommand{\bOmega}{{\boldsymbol \Omega}}

\newcommand{\bS}{{\boldsymbol S}}
\newcommand{\bxi}{{\boldsymbol \xi}}

\newcommand{\bpsi}{{\boldsymbol \psi}}
\newcommand{\C}{{\mathbb C}}

\newcommand{\N}{{\mathbb N}}

\newcommand{\D}{{\partial}}

\begin{document}

\title{Pfaffian form of the \\
Grammian determinant solutions\\ of the
BKP hierarchy}

\author{Q. P. Liu$^1$\thanks{On leave of absence from
Beijing Graduate School, CUMT, Beijing 100083, China}
\thanks{Supported by {\em Beca para estancias temporales
de doctores y tecn\'ologos extranjeros en
Espa\~na: SB95-A01722297}}
   $\,$ and Manuel Ma\~nas$^{2,1}$\thanks{Partially supported by CICYT:
 proyecto PB95--0401}\\
$^1$Departamento de F\'\i sica Te\'orica II,\\ Universidad
Complutense,\\ E28040-Madrid, Spain.\\ $^2$Departamento de
Matem\'atica Aplicada y Estad\'\i stica,\\ Escuela Universitaria de
Ingenier\'\i{}a T\'ecnica Areona\'utica,\\ Universidad
Polit\'ecnica de Madrid,\\ E28040-Madrid, Spain.}
\date{}

\maketitle

\begin{abstract}
The Grammian determinant type solutions of the KP hierarchy,
obtained through  the vectorial binary Darboux transformation, are
reduced, imposing suitable differential constraint on the
transformation data, to Pfaffian solutions of the BKP hierarchy.
\end{abstract}
\newpage
{\bf 1.} Binary Darboux transformations for the KP hierarchy
\cite{oevel} after iteration give Grammian determinant expressions
for new potentials and wave functions. For the BKP hierarchy
\cite{date} instead one finds Pfaffian expressions, see
\cite{hirota} for a bilinear approach, dressing the zero
background, and \cite{nimmo} for direct Darboux transformation
approach (in both papers only the BKP equation was considered, not
the hierarchy); in fact the appearance of Pfaffians was described
in \cite{date}.

As the BKP hierarchy is a reduction of the KP hierarchy
\cite{date}, it is natural to expect a relation between Grammian
and Pfaffian expressions. That is,  the Grammian solutions of the
KP hierarchy reduces to Pfaffian solutions of the BKP hierarchy,
when suitable constraints are impossed in the transformation data.
In this short note we show that this correspondece holds.

The letter is organized as follows. Next, in \S 2 we remind the
reader about some basic facts regarding the KP hierarchy and the
vectorial binary Darboux transformation \cite{lm}, and also about
the BKP hierarchy and how to reduce the mentioned vectorial
transformation to it. Section 3 is devoted to show that these
Grammian expressions reduce to Pfaffians.

{\bf 2.} The KP hierarchy can be formulated as
 the compatibility of the
following linear system:
\begin{equation}\label{linear}
\frac{\D\psi}{\D t_n}=B_n(\psi),\quad n\geq 1,
\end{equation}
where
\[
B_n:=\D^n+
\sum_{m=0}^{n-2}u_{n,m}\D^m,
\]
with $\D:=\dfrac{\D}{\D x}$, $x=t_1$. There  is also an adjoint
linear system
\begin{equation}\label{linear*}
\frac{\D\psi^*}{\D t_n}=-B^*_n(\psi^*),\quad n\geq 1,
\end{equation}
where
\[
B^*_n:=(-1)^n\D^n+
\sum_{m=0}^{n-2}(-1)^m\D^mu_{n,m}.
\]
The functions $u_{n,m}$ are differential polynomials in
$u_{2,0}=:u$.

The vectorial binary Darboux transformation is constructed in terms
of the potential defined by the following compatible equations:
\[
\frac{\D\Omega(\bpsi,\bpsi^*)}{\D t_n}=
-\res(\D^{-1}\bpsi \otimes B^*_n\bpsi^*\D^{-1}), \quad n \geq 1,
\]
where we are working in the ring of pseudodifferential operators in
$\D$,  see for example \cite{oevel}, and assuming that
$\bpsi,\bpsi^*$
 take values in a vectorial spaces $V$ and  $W^*$ (the dual of $W$), respectively.
The first equation is
\[
\D(\Omega(\bpsi,\bpsi^*))=
\bpsi\otimes\bpsi^*,
\]
that for $V=V^*=\C^N$ has the form of a Grammian matrix when
$\bpsi^*=\bpsi^\dagger$.

\newtheorem*{vecto}{\textit{Vectorial Binary Darboux Transformation} \cite{oevel,lm}}

\begin{vecto}
For any pair of  vectorial wave and adjoint wave functions,
$\bxi\in V=\C^N,\bxi^*\in V^*$, respectively,  we construct a new
potential $\tilde u$ and  wave function $\tilde\psi$,
\begin{align*}
\tilde u&=u-\D^2(\ln\det\Omega(\bxi,\bxi^*)),\\
\tilde\psi&=\psi-\Omega(\psi,\bxi^*)\Omega(\bxi,\bxi^*)^{-1}\bxi,\\
&=\frac{1}{|\Omega(\bxi,\bxi^*)|}\begin{vmatrix}\psi
&\Omega(\psi,\bxi^*)\\
\bxi &\Omega(\bxi,\bxi^*)\end{vmatrix}.
\end{align*}
\end{vecto}

The BKP hierarchy is the compatibility of
\[
\frac{\D\psi}{\D t_{2n+1}}=B_{2n+1}(\psi),\quad n\geq 0,
\]
where
\[
B_{2n+1}:=\D^{2n+1}+\sum_{m=0}^{n-2}v_{n,m}\D^{2m+1}.
\]
Being $v_{n,m}$ differential polynomials in $v_{3,1}/3=:v$. Observe
that if in the KP hierarchy one has the constraint
\[
B_{2m+1}^*\D+\D B_{2m+1}=0,\quad m\in\N,
\]
then the odd flows correspond to the BKP hierarchy. Thus, given a
BKP $\Psi(x,t_3,\dotsc)$ wave function one can construct wave and
adjoint wave functions for the odd flows restriction of the KP
hierarchy:
\begin{align*}
\psi(x,0,t_3,0,t_5,\dotsc)&=\Psi(x,t_3,\dotsc),\\
\psi^*(x,0,t_3,0,t_5,\dotsc)&=\D\Psi(x,t_3,\dotsc).
\end{align*}
Moreover, we should have the identity
\[
u(x,0,t_3,0,t_5,\dotsc)=v(x,t_3,\dotsc).
\]

 It can be shown that given transformation data $(\bxi,\bxi^*)$
 such that
\[
 \bxi^*(x,0,t_3,0,t_5,\dotsc)=\D\bxi^\t(x,0,t_3,0,t_5,\dotsc)
 \]
 and choosing the potential satisfying the consistent constraint
 \begin{align*}
 \Omega(\bxi,\bxi^*)+\Omega(\bxi,\bxi^*)^\t&=\bxi\otimes\bxi^\t,\\
\Omega(\psi,\bxi^*)+\Omega(\bxi,\psi^*)^\t&=\psi\bxi^\t,
 \end{align*}
the corresponding vectorial binary Darboux transformation, when
applied to a KP wave function $\psi$ coming from BKP one, is such
that $\tilde\Psi(x,t_3,\dots):=\tilde\psi(x,0,t_3,0,t_5,\dotsc)$ is
a wave function of the BKP hierarchy.

Hence, the vectorial binary Darboux transformation can be
constrained such that it preserves the BKP reduction. Thus, we have
Grammian type determinant solutions of the BKP hierarchy.

Observe that the potential matrices $\Omega:=\Omega(\bxi,\bxi^*)$
and $\Omega(\psi,\bxi^*)$ split into its symmetric and
skew-symmetric parts as follows
\begin{align}
\label{S1}\Omega&=\frac{1}{2}\big[\bxi\otimes\bxi^\t+
S\big],\\
\label{S2}\Omega(\psi,\bxi^*)&=\frac{1}{2}\big[\psi\bxi^\t+
S(\psi,\bxi)^\t\big],
\end{align}
where $S$ and $S(\psi,\bxi)$ are skew-symmetric potentials and for
example satisfy
\begin{align*}
\D S&=\bxi\otimes\D\bxi^\t-(\D \bxi)\otimes \bxi^\t,\\
\D S(\psi,\bxi)&=\psi\D\bxi-(\D \psi) \bxi.
\end{align*}

{\bf 3.} Our aim now is to show that these Grammian expressions can
be written, as a consequence of the constraints, in terms of
Pfaffians. First we shall evaluate the determinant of $\Omega$,
from \eqref{S1} we have
\[
|\Omega|=\frac{1}{2^N}|\bxi\otimes\bxi^\t+S|.
\]
Observe that the $j$-th column of $\bxi\otimes\bxi^\t+S$ is
$\xi_j\bxi+\bS_j$, with
\[
\bS_j=(S_{1,j},\dots,S_{j-1,j},0,
S_{j+1,j},\dots,S_{N,j})^\t.\]
 Hence,
\[
|\Omega|=\frac{1}{2^N}\Big[\sum_{j=1}^N\xi_j|S(j)|+|S|\Big]
\]
with the matrix $S(j)$ obtained from $S$ by replacing the $j$-th
column by $\bxi$. Thus, we have
\[
|\Omega|=\frac{1}{2^N}\bigg[\begin{vmatrix} 0 &-\bxi^\t\\
\bxi & S\end{vmatrix}+|S|\bigg],
\]
and recalling that the determinant of an odd skew-symmetric matrix
vanishes we conclude
\[
|\Omega|=
\begin{cases}
\dfrac{1}{2^N}|S|, & \text{ for $N$ even,}\\
\dfrac{1}{2^N}\begin{vmatrix} 0 &-\bxi^\t\\
\bxi & S\end{vmatrix}, &\text{ for $N$ odd.}
\end{cases}
\]
Using the fact that the determinant of an even skew-symmetric
matrix is the square of a Pfaffian \cite{muir,qft}
\[
|\Omega|=
\begin{cases}
\dfrac{1}{2^N}\Pf(S)^2 & \text{ for $N$ even,}\\
\dfrac{1}{2^N}\Big(\Pf\begin{pmatrix} 0 &-\bxi^\t\\
\bxi & S\end{pmatrix}\Big)^2 &\text{ for $N$ odd.}
\end{cases}
\]

Now, we will compute the Pfaffian expressions for the wave
function. As before, we will distinguish between the even and odd
cases.

For even $N=2k$ we use the first representation for the wave
function:
\[
\tilde\psi=\psi-\Omega(\psi,\bxi^*)\Omega(\bxi,\bxi^*)^{-1}\bxi.
\]
Computing the inverse through the Cramer rule we get
\[
\tilde\psi=\psi-\Omega(\psi,\bxi^*)\cdot\frac{\bOmega}{|\Omega|}
\]
where
\[
\bOmega=
  \begin{pmatrix}
    |\Omega(1)|\\\vdots\\|\Omega(2)|
     \end{pmatrix}
     \]
with $\Omega(j)$ obtained from $\Omega$ by replacing the $j$-th
column by $\bxi$; an alternative formula is
\[
\bOmega=\frac{1}{2^{N-1}}
  \begin{pmatrix}
    |S(1)|\\\vdots\\|S(N)|
     \end{pmatrix}=:\frac{1}{2^{N-1}}\bS.
     \]
 Using \eqref{S2} we find
\[
\tilde\psi=\psi-\Omega(\psi,\bxi^*)\cdot
\frac{\psi\bxi^\t+S(\psi,\bxi)}{2|\Omega|},
\]
and noting that
\[
\bxi\cdot\bOmega=\frac{1}{2^{N-1}}
  \begin{vmatrix}
    0 & -\bxi^\t \\
    \bxi & S
  \end{vmatrix}=0.
\]
we find out
\[
\tilde\psi=\psi-S(\psi,\bxi)\cdot\frac{\bS}{|S|}.
\]
It can be shown \cite{qft}
\[
|S(i)|=\Pf(S)\Pf(S[i])
\]
where $S[i]$ is obtained from $S$ by replacing their
 $i$-th row and $i$-th column with $-\hat\bxi^\t$ and $\hat\bxi$, respectively,
 where
 \[
 \hat\bxi:=
  \begin{pmatrix}
\xi_1     \\\vdots\\\xi_{i-1}\\0
\\\xi_{i+1}\\\vdots\\\xi_N
  \end{pmatrix}.
 \]
Thus,
\[
\tilde\psi=\psi-S(\psi, \bxi)\frac{(\Pf(S[1]),
\cdots,\Pf(S[N]))^\t}{\Pf(S)}=
\frac{
\Pf \begin{pmatrix}     0&\psi & S(\psi, \bxi)^\t\\
                   -\psi & 0   &-\bxi^\t\\
                   -S(\psi,\bxi)&\bxi&S
                   \end{pmatrix}}{\Pf(S)},
\]
where we have used the expansion rule for Pfaffians. This complete
the even case and we proceed with odd case. Now, we use the second
representation for the wave function, namely,
\[
\tilde\psi=\frac{\begin{vmatrix}\psi&\Omega(\psi, \bxi^*)\\
                        \bxi&\Omega\end{vmatrix}}{|\Omega|}.
\]
Notice that
\[
\begin{vmatrix}\psi&\Omega(\psi, \bxi^*)\\
                        \bxi&\Omega\end{vmatrix}
=\frac{1}{2^{N-1}}\begin{vmatrix}\psi&S(\psi,\bxi)^\t\\
\bxi&S
\end{vmatrix},
\]
where we used  \eqref{S1} and \eqref{S2}.

As in the even case, we have
\[
\begin{vmatrix}
\psi&S(\psi,\bxi)^\t\\
\bxi&S\end{vmatrix}
=\Pf\begin{pmatrix} 0&S(\psi,\bxi)^\t\\
                 -S(\psi,\bxi)&S\end{pmatrix}
                 \Pf\begin{pmatrix}0&- \bxi^\t\\
                 \bxi&S\end{pmatrix}.
\]
Using above formula, we arrive
\[
\tilde\psi=\frac{\Pf\begin{pmatrix} 0&S(\psi,\bxi)^\t\\
                 -S(\psi,\bxi)&S\end{pmatrix}}
                 {\Pf\begin{pmatrix}0&-\bxi^\t\\
                 \bxi&S\end{pmatrix}}.
\]

We can summarize the above results as follows

%\begin{Th}
\newtheorem*{vectop}{\textit{
Pfaffians and the Reduced Vectorial Binary Darboux Transformation}}

\begin{vectop}
 The reduction of the Grammian determinant type
solutions of the KP hierarchy to the BKP hierarchy gives the
following Pfaffian expressions:
\begin{enumerate}
\item For $N$ even
\begin{align*}
  \tilde v &= v-2\D^2(\ln\Pf(S)),\\
\tilde\Psi&=\frac{
\Pf \begin{pmatrix}     0&\Psi & S(\Psi, \bxi)^\t\\
                   -\Psi & 0   &-\bxi^\t\\
                   -S(\Psi,\bxi)&\bxi&S
                   \end{pmatrix}}{\Pf(S)}.
\end{align*}
\item For $N$ odd
\begin{align*}
  \tilde v &= v-2\D^2\Big(\ln\Pf\begin{pmatrix} 0 &-\bxi^\t\\
\bxi & S\end{pmatrix}\Big),\\[.2cm]
\tilde\Psi&=\frac{\Pf\begin{pmatrix} 0&S(\Psi,\bxi)^\t\\
                 -S(\Psi,\bxi)&S\end{pmatrix}}
                 {\Pf\begin{pmatrix}0& -\bxi^\t\\
                 \bxi&S\end{pmatrix}}.
\end{align*}

\end{enumerate}

\end{vectop}

\end{document}